\documentclass[journal]{vgtc}                     

\onlineid{0}

\vgtccategory{Research}

\vgtcpapertype{theory/model}

\title{A Design Space for Multiscale Visualization}

\author{
  \authororcid{Mara Solen}{0000-0002-5191-5193},
  \authororcid{Matt Oddo}{0009-0009-3187-4994},
  \authororcid{Tamara Munzner}{0000-0002-3294-3869}
}

\authorfooter{
    \item Mara Solen, Matt Oddo, and Tamara Munzner are with the University of British Columbia, Department of Computer Science. E-mail: \{solen,bofarull,tmm\}@cs.ubc.ca.
}

\abstract{Designing multiscale visualizations, particularly when the ratio between the largest scale and the smallest item is large, can be challenging, and designers have developed many approaches to overcome this challenge. We present a design space for visualization with multiple scales. The design space includes three dimensions, with eight total subdimensions. We demonstrate its descriptive power by using it to code approaches from a corpus we compiled of \change{52} examples, created by a mix of academics and practitioners. We demonstrate descriptive power by analyzing and partitioning these examples into four high-level strategies for designing multiscale visualizations, which are shared approaches with respect to design space dimension choices. We demonstrate generative power by analyzing missed opportunities within the corpus of examples, identified through analysis of the design space, where we note how certain examples could have benefited from different choices. We discuss patterns in the use of different dimension and strategy choices in the different visualization contexts of analysis and presentation. 

Supplemental materials: \url{https://osf.io/wbrdm/}

Design space website: \url{https://marasolen.github.io/multiscale-vis-ds/}}


\teaser{
  \includegraphics[width=\textwidth]{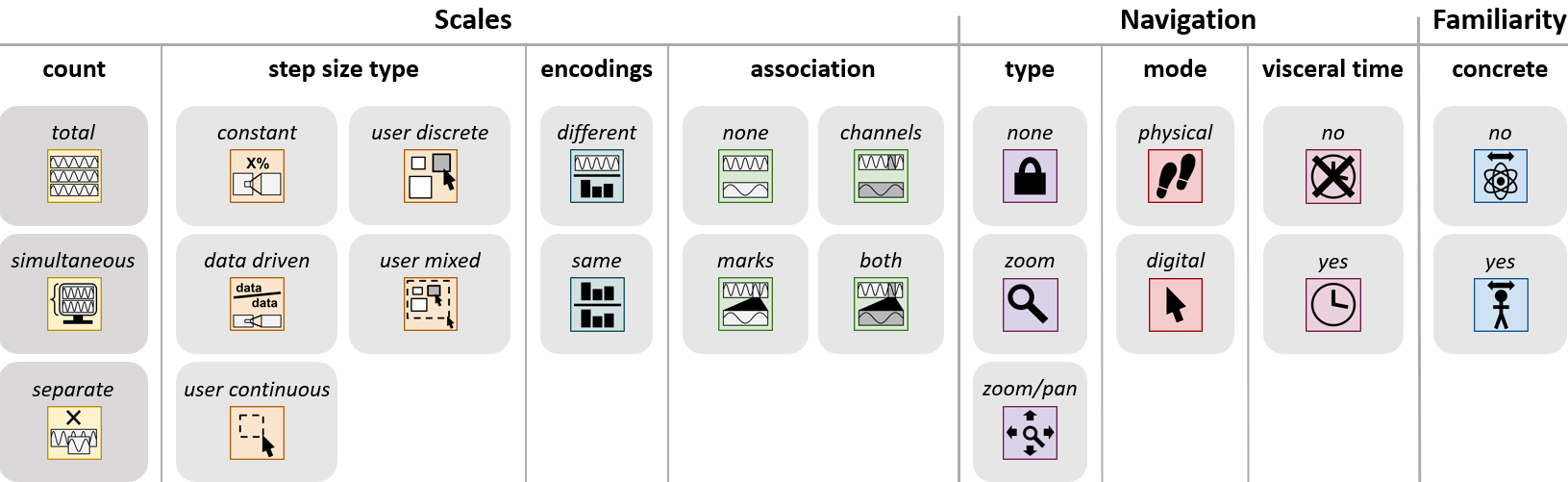}
  \caption{The three dimensions and eight subdimensions of the design space. The quantitative \textbf{count} subdimension is shown with its components. All other subdimensions are categorical and are shown with their choices.}
  \label{fig:dimensions}
}

\graphicspath{{figs/}{figures/}{pictures/}{images/}{./}} 

\usepackage{tabu}                      
\usepackage{booktabs}                  
\usepackage{lipsum}                    
\usepackage{mwe}                       

\usepackage{mathptmx}                  

\usepackage[normalem]{ulem}
\usepackage{subcaption}

\usepackage{soul}
\usepackage{adjustbox}
\usepackage{array}
\usepackage{wrapfig}

\begin{document}

\def\UrlBreaks{\do\/\do-}

\sethlcolor{lightgray}

\newcommand{\tm}[1]{\textcolor{red}{TM: #1}}
\newcommand{\ms}[1]{\textcolor{magenta}{MS: #1}}

\definecolor{ChangedContentColor}{HTML}{0000cd}  
\newenvironment{NEW_ENV}{\color{ChangedContentColor}}{}

\definecolor{DeletedContentColor}{HTML}{C7372F}  

\newcommand{\change}[1]{{#1}}
\newcommand{\delete}[1]{}

\makeatletter
\def\testclr#1#{\@testclr{#1}}
\def\@testclr#1#2{{\fboxsep\z@\fbox{\colorbox#1{#2}{\phantom{XX}}}}}
\newcommand*\Color[1]{\textsl{#1}}
\makeatother

\newcommand\subfigxkcd{A}
\newcommand\subfigmizbee{B}
\newcommand\subfigrivet{C}
\newcommand\subfigmandelbrot{F}
\newcommand\subfigpowersoften{H}
\newcommand\subfigevevis{I}
\newcommand\subfigdelve{E}
\newcommand\subfigmelange{D}
\newcommand\subfiglargevis{F}

\newcommand\iconlines{3}
\newcommand\iconwidthouter{0.07}
\newcommand\iconwidthinner{1.4}


\maketitle
\section{Introduction and Background}
\label{sec:introduction-and-background}

Designing visualizations where the smallest data items can be clearly seen from a high level is challenging. Visualizations typically have a limit to the space they can use, and yet must encode items in the limited space while retaining distinguishability. Also, items may be smaller than can be represented, for example if they require subpixel sizing. 

This challenge arises in a variety of disciplines. Gillmann et al.~include the challenge of designing multiscale visualizations in their ten open challenges in medical visualization \cite{gillmann2021ten}. They comment on how multiple scales are often not visualized simultaneously in medical imaging, and that the field needs to find techniques for integrating the different scales together, suggesting high-level ideas such as focus and context, zooming, and filtering. Similarly, St\aa hlbom et al.~describe that working with multiscale data is a challenge faced by those working on DNA sequencing, as they need to analyze the data at varying levels \cite{staahlbom2022visualization}. Solen et al. describe the challenge of designing a digital exhibit called DeLVE for educating museum visitors about the geological and biological history across varying scales of time, where visitors need to be able to relate the various scales to each other \cite{solen2024delve}.

Visualization designers and researchers have used many techniques to address this challenge, ranging from interactive zooming to multiple simultaneously-visible scales. However, no framework exists to support visualization creators with low-level design decisions while facing this challenge. To meet this need, we present a design space for these scenarios, motivated by the authors' work on and challenges with designing multiscale visualization systems. Rather than exhaustively describing all possible idiom choices for an individual scale, we focus on the larger structure \textit{between} scales, such as the number of scales, how they are related to one another, and how one navigates them. Additionally, we do not consider non-linearity within a single view, such as the use of exponential or logarithmic scales, but we do include non-linearity assembled from multiple linear scales.

Multiscale visualization is used to address two challenges that may arise when visualizing the full dataset on a single linear scale that fits fully within a human range of vision. The first challenge is when the smallest item is not visible, for example if its rendered size results in it being smaller than a pixel, or the smallest manufacturable detail size for physicalizations. The second challenge is when the smallest items are not clearly separable, for example when multiple items must fit within a single pixel or smallest manufacturable detail size. Thus the need for multiscale visualization depends heavily on context; differences in factors like screen pixel density or physical material used to display the visualization will directly impact the need for such solutions.

Our primary contribution is a design space for multiscale visualization, validated via second-stage analysis yielding strategies for designing multiscale visualizations, missed opportunities within the corpus, and dimension choice patterns in different visualization contexts. We also provide a secondary contribution: a coded corpus of \change{52} examples of visualizations developed by both researchers and practitioners.

\section{Related Work}
\label{sec:relwork}

We now discuss related work, first covering work that analyzes collections of multiscale visualizations and then discussing design spaces.

\subsection{Existing Multiscale Frameworks}

Existing work has investigated visualizations to help users understand scales and the use of multiscale visualization. In their work on concrete scales, Chevalier et al.~provide a framework for the varied use of the technique in visualization \cite{chevalier2013using}. While we consider the use of concrete scales as a dimension of our design space, we do not further break its use down into types of concrete scales. The other dimensions are independent from concrete scales.

Garrison et al.~conduct a survey on and construct an overview of visualization in physiology that focuses on multiscale problems \cite{garrison2022trends}\change{, however they}\delete{. While their work covers many visualization examples, they focus on breaking down their findings by different parts of physiology research and} do not provide a model or guidance for visualization design elements.
Jakobsen et al.~investigated the relationship between display size, information space, and scale \cite{jakobsen2013interactive}. \delete{They conduct two user experiments, varying visualization technique, display sizes, and the mapping between the information space and the display size. They report their findings and discuss how the varying factors impact a user's ability to complete a task.} They do not provide lower-level guidance for visualization design as we aim to do.
\change{Tominski et al.~surveyed visualizations with interactive lenses \cite{tominski2017interactive}, but their scope is a subset of ours as lenses are a subset of multiscale visualization.}


\change{Shneiderman's visual information-seeking mantra of overview first, zoom and filter, details on demand \cite{shneiderman2003eyes} intersects the scope of multiscale visualizations, which can be examples of it, but some mantra examples are not multiscale and some multiscale approaches do not use the mantra. Luciani et al.'s mantra of details-first, show context, overview last similarly intersects our scope \cite{luciani2018details}.}  

\change{Information-theoretic visualization \cite{chen2010information} provides a framework for considering uncertainty reduction by showing location and orientation for zooming, which relates to our design space's Association dimension.}

The most closely related work to our own is the structured literature analysis of design practices in multiscale visualization research by Cakmak et al.~\cite{cakmak2021multiscale}. Their work differs from ours, as where they describe the state of research on multiscale visualization, we aim to provide a framework of lower-level design components. Their coding scheme is very different than the dimensions and choices in our own design space, as where they describe high level idiom and interaction choices, we provide lower-level design components which are independent from idiom. Further, their scope is far larger than our own, covering many aspects of multiscale visualization that extend beyond the narrower context of the topic that we investigate. However, their coverage is more narrow, because their analysis covers only the academic literature; they exclude non-academic examples of real-world use by practitioners, which we do include. This difference leads to many of their dimensions and choices to be far outside the scope of our work.

Several of their design considerations do touch on our concerns. The most relevant are \textit{Understand relations across different scales}, \textit{Guide users during multiscale navigations}, \textit{Visualize abstraction measurements across scales}, and \textit{Design tailored multiscale domain visualizations}. Our design space can be seen as a response to these calls for action, providing a structure for analysis to address these very questions. Similarly, our design space responds to one of the open research questions they identified, their call for further quantification of visual scalability. We use their paper as a seed paper for examples from academic literature, and we incorporate the search terms they used in our systematic literature search.

While these existing frameworks support designers and researchers in a variety of ways, none of them explicitly enumerate the possible design choices for constructing multiscale visualizations.

\subsection{Design Spaces}

Design spaces impose systematic structure on a set of possibilities for a specific problem, capturing the key variables at play. They provide an actionable structure for systematically reasoning about solutions~\cite{elliott2021design}. Describing and analyzing portions of a design space allows us to understand differences among designs and suggest new possibilities~\cite{card1997structure}. They increase cognitive efficiency and support inferences, by grouping similar instances together to facilitate reasoning about classes rather than instances~\cite{ralph2019toward}.

Visualization researchers have developed design spaces for a variety of topics. Goffin et al.~created a design space for word-scale visualizations \cite{goffin2014wordscale}, which, similar to our design space, focused on the design components of the visualizations within their scope. \delete{However, it is useful to consider design spaces for concepts outside of design components.} Schulz et al.~use a design space to describe abstract visualization tasks \cite{schulz2013tasks}. Elliot et al.~construct a design space to describe methods, specifically for vision science research on visualizations \cite{elliott2021design}. Kim et al.~provide a design space around accessible visualization which encompasses a combination of these concepts, including design components and abstract tasks as well types of users and technologies \cite{kim2021accessible}. No previous work provides a design space for multiscale visualization; \change{we address} that gap.

\section{Methods}
\label{sec:methods}

The development and validation of our design space took place in three stages: Initialize, Expand, Refine. \change{These three stages reflect the development of the design space; we initially constructed it informally for use in another project, then gradually formalized it as we developed it into a standalone framework.} We iteratively created a corpus of examples to guide the creation and assess the value of the design space, adding new relevant examples and removing those that were no longer in scope at each stage. While collecting examples, we iteratively analyzed the corpus to construct the eight dimensions of the design space. We then analyzed the dimension choices these examples used to construct a set of four strategies, and used the design space choices to identify missed opportunities in certain examples.

\subsection{Corpus Collection and Design Space Construction}

We now discuss the process of each stage of corpus collection and dimension iteration\change{, and the paper's scope}. In the first two stages the example collection was done by two of the authors, and in the final stage the systematic literature search was conducted by the first author. 


\change{For all three stages, we conducted thematic analysis \cite{braun2012thematic} to generate codes. The first author iteratively coded the data, discussing codes with the last author between coding rounds to check for clarity.}

\subsubsection{Initialize}

In the Initialize stage, we constructed an \textbf{initial set} of 21 examples. We began with a small number of examples that we were already aware of or were provided to us by domain expert co-authors with whom we collaborated in other projects. After collecting the initial set, we paused collection to construct an initial version of the design space dimensions, so that we could collect further examples in a later stage to validate those choices. Our initial design space fully described the initial set. 

\subsubsection{Expand}

In the Expand stage, we focused on increasing the coverage of the example corpus to validate our initial design space. We collected 16 more examples, called the \textbf{expansion set}, primarily through forward and backward chaining on our existing academic examples but also from Google searches for practitioner work, additional suggestions, and author memory, resulting in a total of 37 examples in the corpus. The additional examples led to design space refinements that involved both introducing new dimensions to capture more differences and eliminating or merging uninformative dimensions in an effort to improve distinguishability between examples. After modifying the design space, we re-coded the initial set in the new dimensions.

\begin{figure*}
    \centering
    \includegraphics[width=0.96\linewidth]{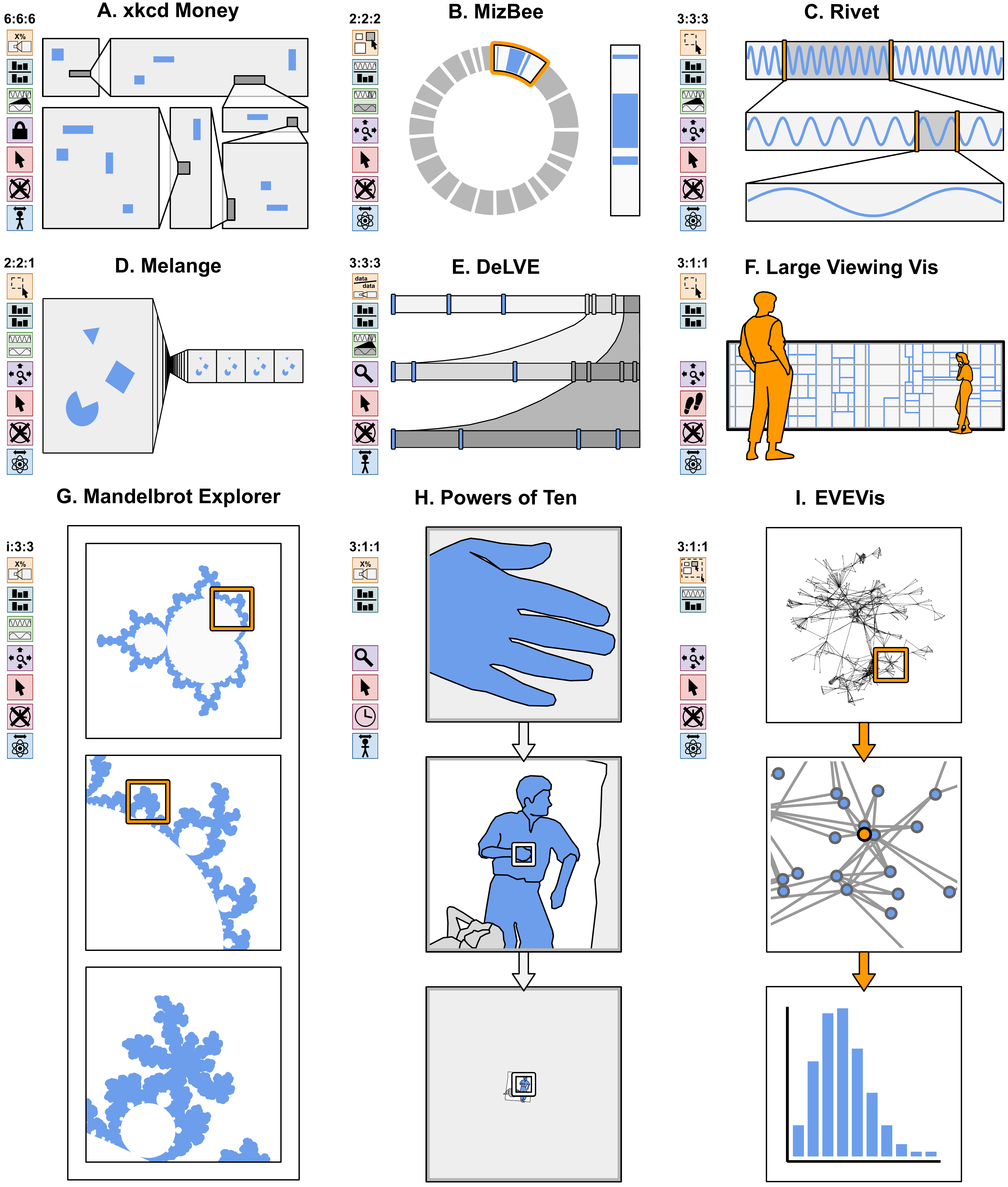}
    \caption{Nine representative examples from the corpus, stylized for clarity. Blue indicates marks or features in the visualization. Orange indicates user interaction. Examples A-G are all on one screen. In examples H and I, the arrows indicate changes to what is shown on screen. \change{When the scale counts here differ from those of the original examples listed in Table~\ref{tab:coded-examples-dimensions}, our stylization involved a change to scale counts for simplicity.} A) xkcd Money \cite{munroe2011xkcdmoney} B) MizBee \cite{meyer2009mizbee}. C) Rivet (multi-tier strip chart) \cite{bosch2000rivet}. D) Melange \cite{elmqvist2008melange}. E) DeLVE \cite{solen2024delve}. F) Large Viewing Vis \cite{isenberg2013largeviewingvis}. G) The Mandelbrot Explorer \cite{bau2009mandelbrot}. H) Powers of Ten \cite{eames1968powersoften}. I) EVEVis \cite{miller2011evevis}.}
    \label{fig:example-superfig}
\end{figure*} 

\subsubsection{Refine}
Since the Expand stage resulted in changes to the design space, we sought out further examples  to validate the newly-refined dimensions. 
By this final Refine stage, we had exhausted all examples from author memory and domain expert suggestion, the majority of which were in real-world use rather than from academic literature. We \change{conducted} a systematic literature search to collect relevant examples from academic literature, validate the design space, and finalize the dimensions. 

We began by reviewing each of the 122 articles included in Cakmak et al.'s collection from 2021 \cite{cakmak2021multiscale}\change{, which led to} 19 new papers and 20 new examples (one paper contributed two systems). 

To collect more recent examples, we then conducted a search for multiscale visualization examples between 2021 and 2025. We used the Bielefeld Academic Search Engine (BASE), which meets quality requirements for academic literature searches \cite{gusenbauer2019academic}, and filtered the search to only include journal and conference articles since 2021 (inclusive). We used the same keywords used by Cakmak et al., namely the keyword ``visualization'' paired with one of the following terms: ``multiple scales'', ``multiresolution'', ``multiple levels of detail'', ``multi-level'', and ``multiscale'' \cite{cakmak2021multiscale}. We chose the same keywords for consistency and because we agree with their previous analysis that these are the most commonly used terms for these techniques. Our goal was to reach theoretical saturation to validate our design space, not to collect a fully exhaustive set of examples of multiscale visualization instantiations. 

Our initial search returned 161 results. We then reviewed each article and used the same inclusion criterion as above, coverage of multiscale visualization systems,
yielding 8 additional articles. \delete{In total, data collection in this stage, including both} \change{The} backward chaining from Cakmak et al.~and the keyword search\delete{,} led to the 28 new examples that we call the \textbf{systematic set}. We coded these new examples in the design space. 

We also removed 14 previously-collected examples as they no longer met our finalized inclusion criteria. \delete{These excluded examples did not showcase multiscale visualization: most of them instead showed large quantities or ranges of data on a single scale, which we had considered in earlier versions of the design space that had slightly broader scope but did not fit into our final version.} \change{We added one additional example after our systematic literature search based on reviewer recommendation.} At the end of the Refine stage, our corpus contained 52 examples: 10 examples remained from the initial set, 13 from the expansion set, and 29 from the systematic set. 

During this iteration, we did not identify any meaningful differences between examples that were not already described by our design space, validating that we had achieved saturation. However, our reflection on the entire corpus at the end of this stage led to refinements to the design space to improve understandability. Finally, we re-coded all \change{52} examples in the corpus into our final set of dimensions and choices. We present the full and final design space in detail in Section~\ref{sec:design-space}.

\subsubsection{Scope}

\change{The scope of this design space includes only visualizations that include multiple geometric scales. We include examples from both real-world use and from academic literature, although we excluded papers that did not concretely illustrate or demonstrate a system or technique. We also exclude examples that only had multiple semantic scales without multiple geometric scales.}

\change{During the first two stages of the project, we had a slightly larger scope which included examples that attempted to help users conceptualize large scales by showing only a small region of a scale and enabling navigation along that scale. These examples relied on the users' sense of visceral time as they navigated along that scale to communicate how big the overall scale was. For simplicity, and to align with well-known terms in the field, we exclude these examples and restrict the scope to multiscale visualization as of the Refine stage.}

\subsection{Analysis of the Coded Corpus}

We also sought to identify implications for design using our design space. To this end, we conducted analysis in three ways. We do not make quantitative claims about usage of our design space's dimensions, as we did not follow an exhaustive example collection method such as GEViT \cite{crisan2019systematic}. Instead, we focus on qualitative analysis methods.

\subsubsection{Strategies}

First, we analyzed coded examples within the design space to find meaningful groups through iterative coding in three steps. For this analysis, the first author \change{conducted thematic analysis \cite{braun2012thematic}}. The author began by coding similar examples into groups that only differed by one or two dimensions. Then, they determined which of the dimensions remained consistent within each group. They used these sets of consistent dimensions to give meaningful names to the groups and define precisely what makes an example fit within a group. They then began this process again, but in the first step would code the groups rather than the individual examples, leading them to merge similar groups into single, larger groups. Once they were no longer able to merge groups together, they applied the definitions of the groups to the entire corpus again to ensure that they had the correct examples in each group and that each example fit into exactly one group each. This process led them to identify four groups, which we call \textbf{strategies}, namely shared approaches with respect to design space choices. We discuss the results of this analysis in Section~\ref{sec:strategies}.

\subsubsection{Missed Opportunities}

We also analyzed missed opportunities in our corpus that we identified through usage of the design space in general, and the interaction between the choices for strategy and for the dimension pertaining to purpose. 
For this analysis, the first author analyzed the corpus for missed opportunities, and finalized them in discussion with the last author.
We noted where there were large discrepancies such as heavily used or underused choices across the entire corpus and within specific strategies. We identified examples using the discrepancies and analyzed them with respect to whether using alternative choices could have improved the design. We chose to undertake in-depth redesigns for two examples. We discuss our findings from this analysis in Section~\ref{sec:missed-opportunities}. 

\subsubsection{Visualization Context: Analysis vs. Presentation}

The first author additionally analyzed our corpus with respect to the context in which they were used: analysis, gleaning new information, or presentation, communicating already-gleaned information. 
We coded the corpus by their context, with some left blank as they described techniques rather than deployed visualizations. We then analyzed these two groups for differences in distribution in design choice, strategy, and source, for example looking for choices used often in one context but rarely in the other. We discuss our findings in Section~\ref{sec:visualization-context}.

\section{The Design Space}
\label{sec:design-space}

The design space contains 3 independent dimensions, each with between one and four subdimensions, for a total of eight subdimensions. Figure~\ref{fig:dimensions} shows an overview of the dimension hierarchy. In this section, we describe each dimension in detail (bold), and explain the choices within it (italics) by referring to examples in the corpus. We also provide nine representative examples that are simplified and stylized evocations of examples found within the corpus, shown in Figure~\ref{fig:example-superfig}. We chose these examples to, together, cover all concepts described in the design space's dimensions, enabling us to describe them more clearly. Table~\ref{tab:coded-examples-dimensions} shows all examples and their codes for each dimension. The coded design space is additionally available as an interactive website with filters at~\url{marasolen.github.io/multiscale-vis-ds/}.

\begin{table*}[!ht]
    \centering
    \small
    \begin{tabular}{p{0.14\linewidth}   p{0.04\linewidth}   p{0.05\linewidth}   p{0.02\linewidth}   p{0.04\linewidth}   |   >{\raggedleft}p{0.03\linewidth}   p{0.07\linewidth}   p{0.06\linewidth}   p{0.04\linewidth}   |   p{0.03\linewidth}   p{0.05\linewidth}   p{0.04\linewidth}   |   p{0.06\linewidth} }
         &  &  &  &  &  \multicolumn{4}{c|}{\textbf{Scales}} & \multicolumn{3}{c|}{\textbf{Navigation}} & \textbf{Familiarity} \\ \hline
        \textbf{Example (short title)} & \textbf{Citation} & \textbf{Source} & \textbf{Use} & \textbf{Stage} & \textbf{count} & \textbf{step type} & \textbf{encodings} & \textbf{assoc.} & \textbf{type} & \textbf{mode} & \textbf{visceral time} & \textbf{concrete}  \\ \hline \hline
        
        \multicolumn{12}{c}{\textbf{Single-View Pan and Zoom}} \\ \hline
        
        Zoom Line Chart & \cite{tiwari2020zoomlinechart}                          & practitioner & \change{T} & init.  & 3:1:1    & user cont.  & same      &          & z/p    & digital  & no  & no    \\ \hline
        Cuttlefish (fig 6) & \cite{waldin2019cuttlefish}                          & academic     & \change{T} & expand & 3:1:1    & user cont.  & same      &          & z/p    & digital  & no  & no    \\ \hline
        EVEVis & \cite{miller2011evevis}                                          & academic     & A & expand & 4:1:1    & user mixed  & different &          & z/p    & digital  & no  & no    \\ \hline
        Multilevel Poetry & \cite{mittmann2016multilevelpoetry}                   & academic     & A & expand & 4:1:1    & user disc.  & different &          & z/p    & digital  & no  & no    \\ \hline
        Multiscale Trace & \cite{ezzati2014multiscaletrace}                       & academic     & A & expand & 5:1:1    & user cont.  & different &          & z/p    & digital  & no  & no    \\ \hline
        Europe OSM & \cite{zacharopoulou2021europeosm}                            & academic     & A & refine & 3:1:1    & user cont.  & different &          & z/p    & digital  & no  & no    \\ \hline
        Zoomable Treemaps & \cite{blanch2007zoomabletreemaps}                     & academic     & \change{T} & refine & 3:1:1    & user disc.  & same      &          & z/p    & digital  & no  & no    \\ \hline
        Chameleon & \cite{waldin2016chameleon}                                    & academic     & \change{T} & refine & 5:1:1    & user cont.  & same      &          & z/p    & digital  & no  & no    \\ \hline
        Hierarchical Route Maps & \cite{wang2014hierarchicalroutemaps}            & academic     & A & refine & 3:1:1    & user cont.  & same      &          & z/p    & digital  & no  & no    \\ \hline
        Large Viewing Vis & \cite{isenberg2013largeviewingvis}                    & academic     & P & refine & 3:1:1    & user cont.  & same      &          & z/p    & physical & no  & no    \\ \hline
        Kyrix-S & \cite{tao2020kyrix}                                             & academic     & \change{T} & refine & 15:1:1   & user cont.  & same      &          & z/p    & digital  & no  & no    \\ \hline
        Membrane Mapping & \cite{waltemate2014membranemapping}                    & academic     & A & refine & 4:1:1    & user cont.  & same      &          & z/p    & digital  & no  & no    \\ \hline
        Execution Trace Vis & \cite{trumper2013executiontracevis}                 & academic     & A & refine & 2:1:1    & user cont.  & same      &          & z/p    & digital  & no  & no    \\ \hline
        MuSE & \cite{furnas1998muse}                                              & academic     & \change{T} & refine & 3:1:1    & user cont.  & same      &          & z/p    & digital  & no  & no    \\ \hline
        ScaleTrotter & \cite{halladjian2019scaletrotter}                          & academic     & A & refine & 7:1:1    & user cont.  & same      &          & z/p    & digital  & no  & no    \\ \hline
        SpaceFold & \cite{butscher2014spacefoldphysicslenses}                     & academic     & \change{T} & refine & 2:1:1    & user cont.  & same      &          & z/p    & digital  & no  & no    \\ \hline
        TagNetLens & \cite{gou2010tagnetlens}                                     & academic     & A & refine & 2:1:1    & user disc.  & same      &          & z/p    & digital  & no  & no    \\ \hline
        Hierarchy Vis & \cite{holten2008hierarchyvis}                             & academic     & \change{T} & refine & 2:1:1    & user disc.  & same      &          & z/p    & digital  & no  & no    \\ \hline
        Chemical Vis & \cite{yamazawa2008chemicalvis}                             & academic     & A & refine & 2:1:1    & user cont.  & different &          & z/p    & digital  & no  & no    \\ \hline
        MissVis & \cite{alsufyani2024multi}                                       & academic     & A & refine & 2:1:1    & user cont.  & same      &          & z/p    & digital  & no  & no    \\ \hline
        
        \multicolumn{12}{c}{\textbf{Simultaneous Occluding Embed}} \\ \hline
        
        Melange & \cite{elmqvist2008melange}                                      & academic     & \change{T} & init.  & 2:2:1    & user cont.  & same      & none     & z/p    & digital  & no  & no    \\ \hline
        FingerGlass & \cite{kaser2011fingerglass}                                 & academic     & \change{T} & expand & 2:2:1    & user cont.  & same      & none     & z/p    & digital  & no  & no    \\ \hline
        Tabletop Gestures & \cite{rusnak2018tabletopgestures}                     & academic     & \change{T} & refine & 2:1:1    & user cont.  & same      & marks    & z/p    & digital  & no  & no    \\ \hline
        Gimlenses & \cite{pindat2013gimlenses}                                    & academic     & A & refine & 3:3:1    & user cont.  & same      & marks    & z/p    & digital  & no  & no    \\ \hline
        GrouseFlocks & \cite{archambault2008grouseflocks}                         & academic     & A & refine & 2:2:\change{1}    & user cont.  & same      & marks    & z/p    & digital  & no  & no    \\ \hline
        Digital Earth & \cite{sherlock2021digitalearth}                           & academic     & A & refine & 3:3:1    & user cont.  & same      & marks    & z/p    & digital  & no  & no    \\ \hline
        AdvEx & \cite{you2023advex}                                               & academic     & P & refine & 2:2:1    & user disc.  & same      & marks    & z/p    & digital  & no  & no    \\ \hline
        Scalable Insets & \cite{lekschas2019scalableinsets}                       & academic     & A & refine & 2:2:1    & user cont.  & same      & marks    & z/p    & digital  & no  & no    \\ \hline
        Multi-Foci COVID Vis & \cite{mactavish2022multifocicovidvis}              & academic     & P & refine & 2:2:1    & user cont.  & different & marks    & z/p    & digital  & no  & no    \\ \hline
        PhysicLenses & \cite{butscher2014spacefoldphysicslenses}                  & academic     & \change{T} & refine & 2:2:1    & user cont.  & same      & marks    & z/p    & digital  & no  & no    \\ \hline
        TissUUmaps & \cite{pielawski2023tissuumaps}                               & academic     & A & refine & 2:2:1    & user cont.  & same      & marks    & z/p    & digital  & no  & no    \\ \hline
        TrailMap & \cite{zhao2013trailmap}                                        & academic     & A & refine & 2:2:1    & user cont.  & same      & none     & z/p    & digital  & no  & no    \\ \hline
        \change{Tangible Views} & \cite{spindler2010tangible}                     & \change{academic}     & \change{P} & \change{refine} & \change{2:2:1}    & \change{user cont.}  & \change{same}      & \change{none}     & \change{z/p}    & \change{physical} & \change{no}  & \change{no}    \\ \hline
        
        \multicolumn{12}{c}{\textbf{Simultaneous Separate Multilevel}} \\ \hline
        
        Multiscale Unfolding & \cite{halladjian2021multiscaleunfolding}           & academic     & A & init.  & 6:6:6    & user cont.  & same      & none     & z/p    & digital  & no  & no    \\ \hline
        Rivet (MTSC) & \cite{bosch2000rivet}                                      & academic     & A & init.  & 6:3:3    & user cont.  & same      & marks    & z/p    & digital  & no  & no    \\ \hline
        Temp Earth & \cite{bredenberg2012tempearth}                               & practitioner & P & init.  & 7:7:7    & data driven & same      & none     & none   &          & no  & no    \\ \hline
        TraXplorer (fig 2) & \cite{javed2010traxplorer}                           & academic     & \change{T} & init.  & 5:5:5    & user cont.  & same      & channels & z/p    & digital  & no  & no    \\ \hline
        Mandelbrot Explorer & \cite{bau2009mandelbrot}                            & practitioner & A & expand & inf:6:6  & constant    & same      & none     & z/p    & digital  & no  & no    \\ \hline
        MizBee & \cite{meyer2009mizbee}                                           & academic     & A & expand & 3:3:3    & user disc.  & different & channels & z/p    & digital  & no  & no    \\ \hline
        PolyZoom & \cite{javed2012polyzoom}                                       & academic     & \change{T} & expand & 4:3:3    & user cont.  & same      & marks    & z/p    & digital  & no  & no    \\ \hline
        Chromoscope & \cite{lyi2023chromoscope}                                   & academic     & A & refine & 3:2:2    & user cont.  & different & none     & z/p    & digital  & no  & no    \\ \hline
        TimeNotes & \cite{walker2015timenotes}                                    & academic     & A & refine & 4:4:4    & user cont.  & same      & both     & z/p    & digital  & no  & no    \\ \hline
        Ragas & \cite{balaji2024ragas}                                            & academic     & A & refine & 2:2:2    & data driven & same      & both     & none   &          & no  & no    \\ \hline
        
        \multicolumn{12}{c}{\textbf{Familiar Zoom}} \\ \hline
        
        DeLVE & \cite{solen2024delve}                                             & academic     & P & init.  & 10:10:10 & data driven & same      & both     & zoom   & digital  & no  & yes   \\ \hline
        Here is Today & \cite{twyman2020hereistoday}                              & practitioner & P & init.  & 11:1:1   & data driven & same      &          & zoom   & digital  & no  & yes   \\ \hline
        Powers of Ten & \cite{eames1968powersoften}                               & practitioner & P & init.  & 40:1:1   & constant    & same      &          & zoom   & digital  & yes & yes   \\ \hline
        xkcd Money & \cite{munroe2011xkcdmoney}                                   & practitioner & P & init.  & 5:5:5    & constant    & same      & marks    & none   & digital  & no  & yes   \\ \hline
        Cell Size and Scale & \cite{utah2014cellsizeandscale}                     & practitioner & P & expand & 9:1:1    & user cont.  & same      &          & zoom   & digital  & no  & yes   \\ \hline
        Scale of the Universe 2 & \cite{huang2012scaleoftheuniverse2}             & practitioner & P & expand & 63:1:1   & user cont.  & same      &          & zoom   & digital  & no  & yes   \\ \hline
        The Size of Space & \cite{agarwal2019thesizeofspace}                      & practitioner & P & expand & 35:1:1   & data driven & same      &          & zoom   & digital  & no  & yes   \\ \hline
        Universcale & \cite{nikon2023universcale}                                 & practitioner & P & expand & 42:1:1   & user cont.  & same      &          & zoom   & digital  & no  & yes   \\ \hline
        US Debt & \cite{godfrey2022usdebt}                                        & practitioner & P & expand & 9:3:3    & data driven & same      & none     & zoom   & digital  & no  & yes   \\ \hline
    \end{tabular}
    \caption{The \change{52} corpus examples grouped according to the 4 strategies, coded by the hierarchical dimensions of the design space. The \textbf{Count} subdimension uses the format \texttt{total:simultaneous:separate} to display its three components. Abbreviations: P~= presentation; A~= analysis; \change{T~= technique that does not specify whether it is intended for use in presentation or analysis contexts;} init.~= initialize; assoc.~= association; cont.~= continuous; disc.~= discrete; inf.~= infinite. z/p~= zoom/pan.}
    \label{tab:coded-examples-dimensions}
    \medskip
\end{table*}

\subsection{Scales}

The \textbf{Scales} dimension describes a visualization's encodings in terms of the number of different scales and how any different scales differ in mapping, whether they share low-level encoding choices, and how they are associated with each other. It includes four subdimensions: \textbf{count}, \textbf{step type}, \textbf{encodings}, \textbf{association}.

\subsubsection{Count}

The \textbf{count} subdimension includes three quantitative components: \textit{total}, \textit{simultaneous}, and \textit{separate}.

\begin{wrapfigure}[\iconlines]{l}{\iconwidthouter\linewidth}
  \centering
  \vspace{-10pt}
  \includegraphics[width=\iconwidthinner\linewidth]{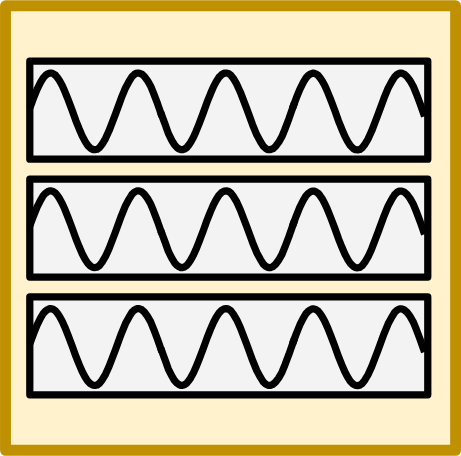}
\end{wrapfigure}
\noindent{\textit{Total.}} The \textit{total} component represents the total number of scales accessible in a visualization. In the case that the number of possible unique scales is discrete, it is straightforward to count. For example, in our simplified version of xkcd's Money webcomic (xkcd Money) \cite{munroe2011xkcdmoney}, a unit chart of different quantities of money on different scales illustrated in Figure~\ref{fig:example-superfig}\subfigxkcd, we can simply count the six scales. \change{The original webcomic has a \textit{total} count of five scales.}

In the case of multiple scales being chosen from a continuous scale, we count the orders of magnitude using the expression \(round(log_{10}(max) - log_{10}(min) + 1)\). In the Powers of Ten example \cite{eames1968powersoften}, a video documentary that gradually zooms between different scales illustrated in Figure~\ref{fig:example-superfig}\subfigpowersoften, the visualization automatically zooms out from an image of two people on a picnic blanket. In our simplified illustration of this example, it zooms from 1 meter to 100 meters, so the \textit{total} component is \change{\(round(log_{10}(100) - log_{10}(1) + 1)=2-0+1=3\)}. \change{The original documentary has 40 \textit{total} scales.}

The Mandelbrot Explorer example \cite{bau2009mandelbrot}, a web page where users can explore Mandelbrot sets by zooming in to areas of their choosing illustrated in Figure~\ref{fig:example-superfig}\subfigmandelbrot, stands out from other examples as the user can zoom an infinite number of times due to its recursive nature. We code \change{the original} example as having an infinite \textit{total} count. 

\begin{wrapfigure}[\iconlines]{l}{\iconwidthouter\linewidth}
  \centering
  \includegraphics[width=\iconwidthinner\linewidth]{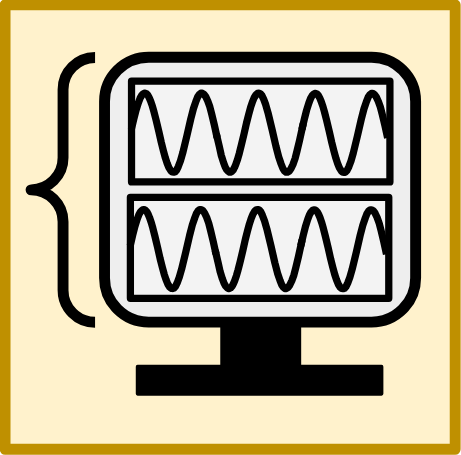}
\end{wrapfigure}
\noindent{\textit{Simultaneous.}} The \textit{simultaneous} component represents the number of different scales that are visible from a viewpoint or on a screen at once. We define visible as where the user can still see the finest detail, whether that is an individual item, the smallest trend, or something else. Powers of Ten only shows a single scale at a time, and the user \change{must} wait for it to zoom out to \change{see} a different scale. 

In our simplified version of MizBee \cite{meyer2009mizbee}, a visualization tool for analysing genomic data at multiple scales illustrated in Figure~\ref{fig:example-superfig}\subfigmizbee, there are two different scales on screen at once. \change{In the original version of MizBee, there are three different scales on screen at once.}

The number of \textit{simultaneous} scales cannot be larger than the \textit{total} component, as any scales which are different and visualized simultaneously must be counted towards the \textit{total} count.

\begin{wrapfigure}[\iconlines]{l}{\iconwidthouter\linewidth}
  \centering
  \vspace{-10pt}
  \includegraphics[width=\iconwidthinner\linewidth]{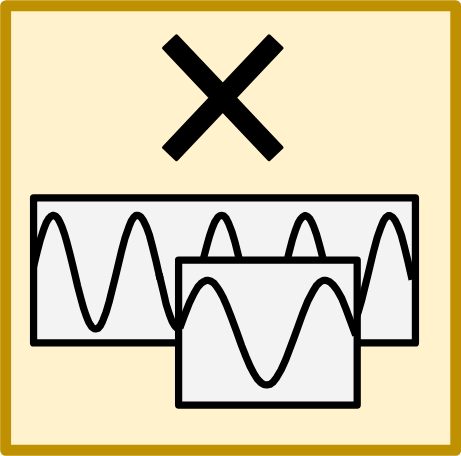}
\end{wrapfigure}
\noindent{\textit{Separate.}} The \textit{separate} component represents the number of simultaneously visible scales that do not occlude each other in any way. In Melange \cite{elmqvist2008melange}, which is a technique that folds a visualization to make the zoomed-in area appear physically closer to the user illustrated in Figure~\ref{fig:example-superfig}\subfigmelange, the zoomed area occludes its zoomed-out range and its neighbouring areas. 
Our representative example Rivet \cite{bosch2000rivet}, which includes a multitier strip chart (MTSC) that shows multiple zoomed levels of computer systems data illustrated as a line chart for simplicity in Figure~\ref{fig:example-superfig}\subfigrivet, shows three separate scales at once without occlusion, so its \textit{separate} component is three.
The number of \textit{separate} scales cannot be larger than the \textit{simultaneous} component, as any scales which are different and visualized simultaneously without occlusion must \change{also count as} \delete{the} \textit{simultaneous}\delete{ count}.

\subsubsection{Step Size Type}

\textbf{Step size type} is a subdimension that describes the relationship between steps from scale to scale in multiscale examples, and it has five choices: \textit{constant}, \textit{data driven}, \textit{user continuous}, \textit{user discrete}, and \textit{user mixed}. In order for a visualization to have steps between scales, it must have multiple scales, so examples with a \textit{total} count of one were left with blank \textbf{step size type} cells.

\begin{wrapfigure}[\iconlines]{l}{\iconwidthouter\linewidth}
  \centering
  \vspace{-10pt}
  \includegraphics[width=\iconwidthinner\linewidth]{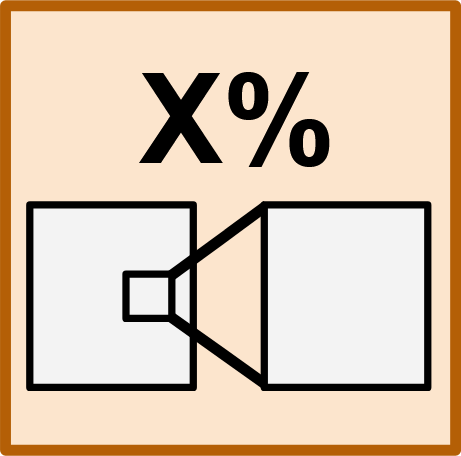}
\end{wrapfigure}
\noindent{\textit{Constant.}} When this dimension is \textit{constant}, it means that the difference between scales, which is typically multiplicative, does not change. The Powers of Ten example zooms out at a constant rate, so the step size is \textit{constant}. xkcd Money with multiplicative increases of 1000 between each scale.

\begin{wrapfigure}[\iconlines]{l}{\iconwidthouter\linewidth}
  \centering
  \vspace{-10pt}
  \includegraphics[width=\iconwidthinner\linewidth]{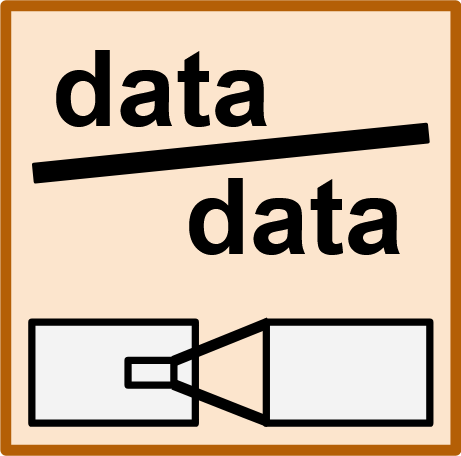}
\end{wrapfigure}
\noindent{\textit{Data driven.}} The \textbf{step size type} being \textit{data driven} means that the scales' mappings change by a data-based amount. DeLVE \cite{solen2024delve}, a visualization technique for showing past events on multiple scales with lines between them to show how the scales relate, is illustrated in Figure~\ref{fig:example-superfig}\subfigdelve. In DeLVE, step sizes are determined by the events in the dataset rather than a constant value or user choice.

\begin{wrapfigure}[\iconlines]{l}{\iconwidthouter\linewidth}
  \centering
  \vspace{-10pt}
  \includegraphics[width=\iconwidthinner\linewidth]{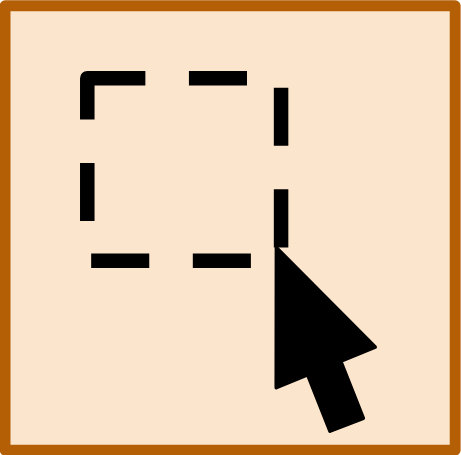}
\end{wrapfigure}
\noindent{\textit{User continuous.}} The \textit{user continuous} option describes when the user chooses the step size from a range. Rivet uses this choice as the user decides on the specific zoom level themself by choosing a range to magnify in the next scale.

\begin{wrapfigure}[\iconlines]{l}{\iconwidthouter\linewidth}
  \centering
  \vspace{-10pt}
  \includegraphics[width=\iconwidthinner\linewidth]{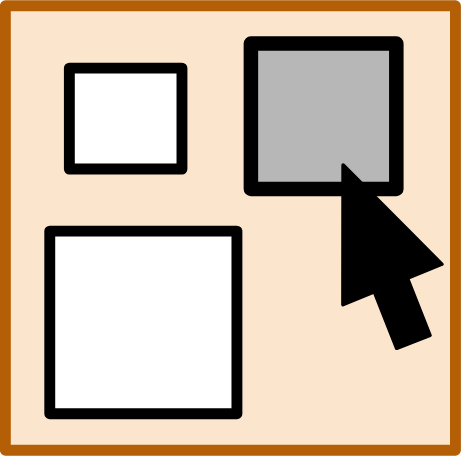}
\end{wrapfigure}
\noindent{\textit{User discrete.}} In some visualizations, designers allow users to choose how to navigate from a discrete set of options, a choice we call \textit{user discrete}. MizBee uses this choice as the user chooses the step size by selecting an area to zoom in to from a set of discrete options. 

\begin{wrapfigure}[\iconlines]{l}{\iconwidthouter\linewidth}
  \centering
  \vspace{-10pt}
  \includegraphics[width=\iconwidthinner\linewidth]{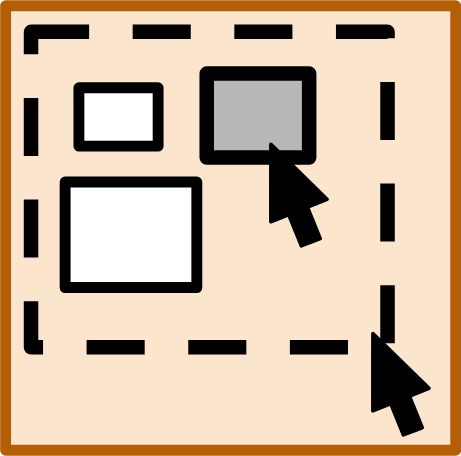}
\end{wrapfigure}
\noindent{\textit{User mixed.}} Finally, some visualizations use both \textit{user continuous} and \textit{user discrete}, which together form the final option for this dimension: \textit{user mixed}. EVEVis ~\cite{miller2011evevis}, a visualization for evolution data at multiple scales illustrated in a stylized \delete{and simplified} manner in Figure~\ref{fig:example-superfig}\subfigevevis, uses \textit{user mixed} as the user can first select a region from a continuous space, then further zooming is done via discrete selections.

\subsubsection{Encodings}

\textbf{Encodings} is a subdimension that describes whether different scales use the same or different visual encodings. This subdimension relies on a visualization using more than one scale, so it is left blank for corpus examples with a \textit{total} of one.

\begin{wrapfigure}[\iconlines]{l}{\iconwidthouter\linewidth}
  \centering
  \vspace{-10pt}
  \includegraphics[width=\iconwidthinner\linewidth]{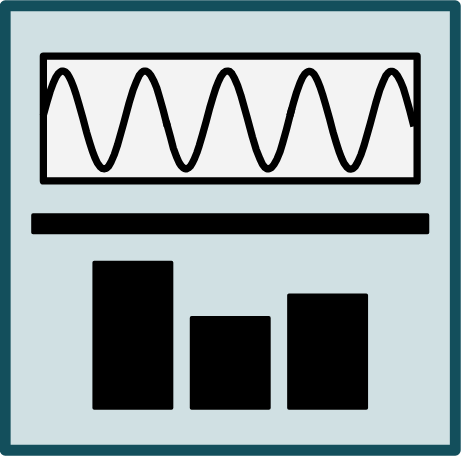}
\end{wrapfigure}
\noindent{\textit{Different.}} Our stylized illustration of EVEVis uses \textit{different} encodings on each scale, with a node link diagram in the upper two levels, the highest of which is the first to be shown, and a bar chart in the lowest level, which is navigated to \change{by the user}. \change{We count any encoding change between scales in an example as \textit{different}.}

\begin{wrapfigure}[\iconlines]{l}{\iconwidthouter\linewidth}
  \centering
  \vspace{-10pt}
  \includegraphics[width=\iconwidthinner\linewidth]{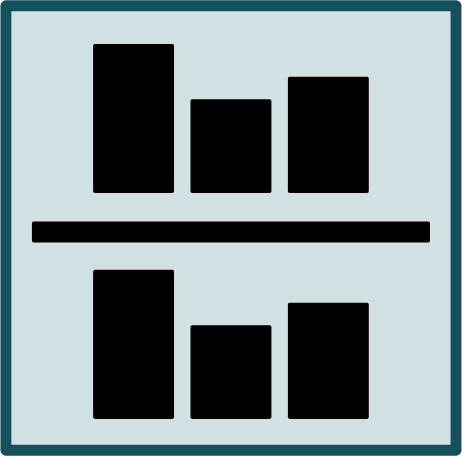}
\end{wrapfigure}
\noindent{\textit{Same.}} In contrast, an example that uses the \textit{same} encoding on every scale is DeLVE. While each scale in DeLVE uses different colours and encodes mostly different data, the encoding choices to use event markers on a timeline remains the same.

\subsubsection{Association}

\textbf{Association} is a subdimension that describes how marks representing the same item can be visually linked across simultaneously visible scales. It relies on a visualization including multiple scales visible at once, and was left blank for examples with \change{one \textit{simultaneous} scale}.

\begin{wrapfigure}[\iconlines]{l}{\iconwidthouter\linewidth}
  \centering
  \vspace{-10pt}
  \includegraphics[width=\iconwidthinner\linewidth]{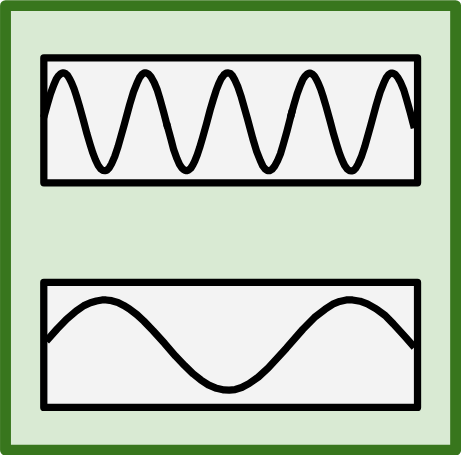}
\end{wrapfigure}
\noindent{\textit{None.}} Mandelbrot Explorer does not use any marks or channels to show association between its different scales. Instead, it simply shows its scales on screen in a grid pattern. We code this subdimension as \textit{none}. \change{If an example uses animation or interaction to show two consecutive scales, we code it as \textit{none}; we differentiate between explicit visual association and a reliance on memory.}

\begin{wrapfigure}[\iconlines]{l}{\iconwidthouter\linewidth}
  \centering
  \vspace{-10pt}
  \includegraphics[width=\iconwidthinner\linewidth]{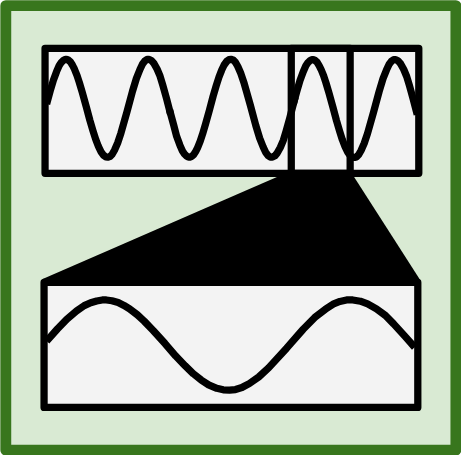}
\end{wrapfigure}
\noindent{\textit{Marks.}} Association can be done by \textit{marks}, often using connection \textit{marks} to show association. An example of this approach is Rivet, where there are lines showing how the lower levels and the focused point of the higher level are associated. 

\begin{wrapfigure}[\iconlines]{l}{\iconwidthouter\linewidth}
  \centering
  \vspace{-10pt}
  \includegraphics[width=\iconwidthinner\linewidth]{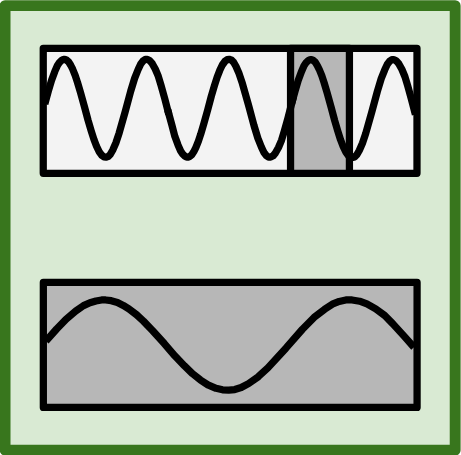}
\end{wrapfigure}
\noindent{\textit{Channels.}} The other option we identified in our corpus was association by visual \textit{channels} such as colour. MizBee uses the colour \textit{channel} to show that the lower-level scale, in blue, is the zoomed in version of the blue part of the higher-level scale.

\begin{wrapfigure}[\iconlines]{l}{\iconwidthouter\linewidth}
  \centering
  \vspace{-10pt}
  \includegraphics[width=\iconwidthinner\linewidth]{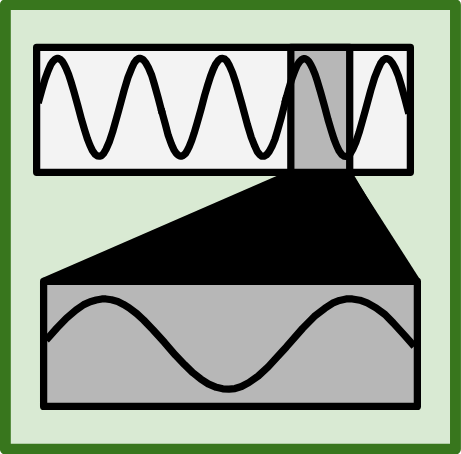}
\end{wrapfigure}
\noindent{\textit{Both.}} DeLVE uses \textit{both} the colour \textit{channel} to show that the lowest-level scale is the zoomed in version of the darkest part of the higher-level scales and line \textit{marks} between scales to show \textbf{association}.

\subsection{Navigation}

The \textbf{Navigation} dimension covers the interaction capabilities of the design\change{, with} three subdimensions: \textbf{type}, \textbf{mode}, and \textbf{visceral time}.

\subsubsection{Type}

\textbf{Type} is a subdimension that describes the ways in which users can navigate between and within scales. 

\begin{wrapfigure}[\iconlines]{l}{\iconwidthouter\linewidth}
  \centering
  \vspace{-10pt}
  \includegraphics[width=\iconwidthinner\linewidth]{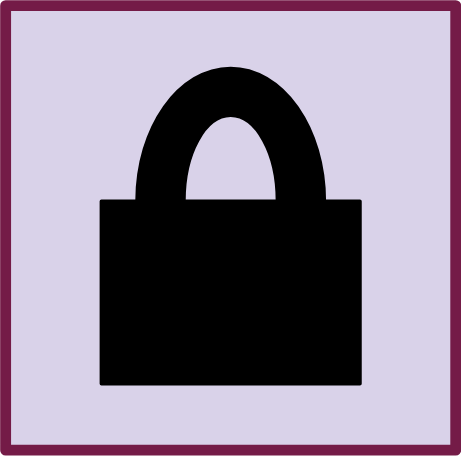}
\end{wrapfigure}
\noindent{\textit{None.}} Some visualizations, such as xkcd Money, are \textit{none}, meaning they do not have any interaction. Although in practice users may use the zoom and pan features of an image browser to inspect its details, there is no explicit navigation support within the visualization itself.

\begin{wrapfigure}[\iconlines]{l}{\iconwidthouter\linewidth}
  \centering
  \vspace{-10pt}
  \includegraphics[width=\iconwidthinner\linewidth]{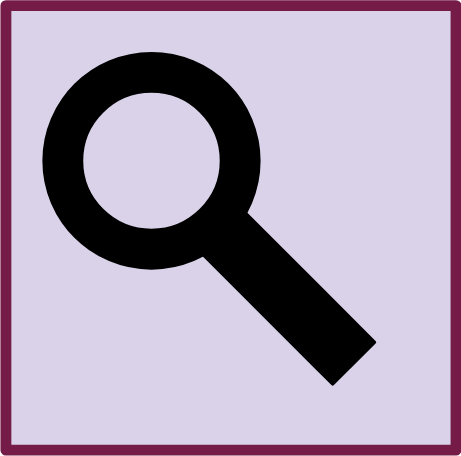}
\end{wrapfigure}
\noindent{\textit{Zooming.}} \textit{Zooming} is navigation that changes the scale or adds a different scale without changing the point of focus. Powers of Ten uses \textit{zooming} to gradually change the scale. Zooming can happen automatically, such as in Powers of Ten, or it can be facilitated by digital means, such as scrolling a mouse wheel, or physical means, such as walking closer or further away.

\begin{wrapfigure}[\iconlines]{l}{\iconwidthouter\linewidth}
  \centering
  \vspace{-10pt}
  \includegraphics[width=\iconwidthinner\linewidth]{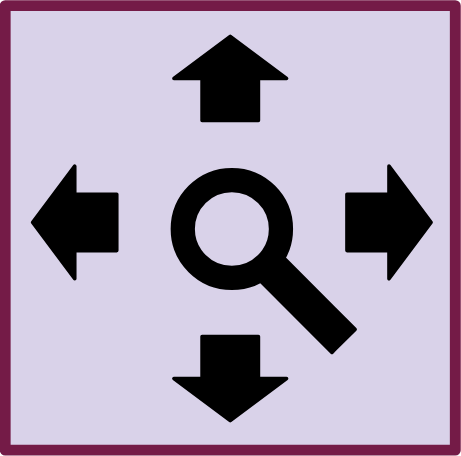}
\end{wrapfigure}
\noindent{\textit{Zooming/Panning.}} Some visualizations additionally incorporate panning, which is navigation that changes the point of focus without changing the scale, as well as zooming. We code these examples as using \textit{zoom/pan}. Rivet allows the user to select a point of focus to zoom in on, effectively panning and zooming at once.

\subsubsection{Mode}

\textbf{Mode} is a subdimension that describes how navigation is done, whether through physically moving oneself or by digitally navigating via computer input devices. If there is no navigation in an example, we left the \textbf{mode} cell blank.

\begin{wrapfigure}[\iconlines]{l}{\iconwidthouter\linewidth}
  \centering
  \vspace{-10pt}
  \includegraphics[width=\iconwidthinner\linewidth]{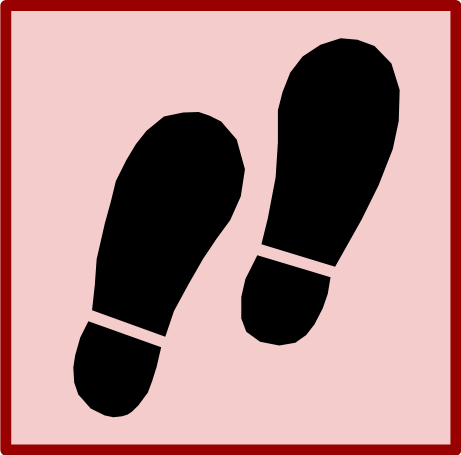}
\end{wrapfigure}
\noindent{\textit{Physical.}} In some navigable visualizations, users control the navigation by physically moving their body. Large Viewing Vis \cite{isenberg2013largeviewingvis}, a technique that uses a large screen with visible features at many levels and that requires users to move closer to to see smaller features illustrated in Figure~\ref{fig:example-superfig}\subfiglargevis, has users both \textit{pan} and \textit{zoom} physically. \change{Tangible Views, where a user physically moves a tablet-like device around a separate, larger screen are another example of the \textit{physical} \textbf{Mode} \cite{spindler2010tangible}.}

\begin{wrapfigure}[\iconlines]{l}{\iconwidthouter\linewidth}
  \centering
  \vspace{-10pt}
  \includegraphics[width=\iconwidthinner\linewidth]{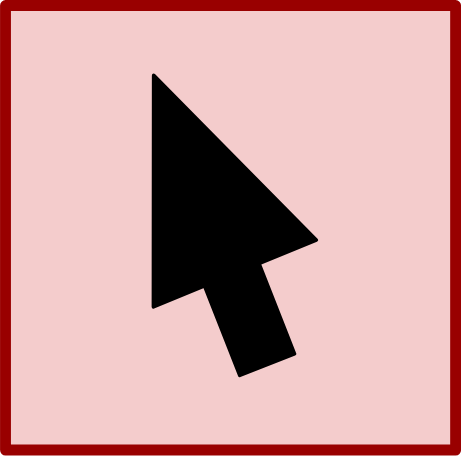}
\end{wrapfigure}
\noindent{\textit{Digital.}} Virtual visualizations are controlled by an input device such as computer mice, keyboards, or touch screens. Users of Mandelbrot Explorer use the mouse to select a region to zoom in to next.

\subsubsection{Visceral Time}

The \textbf{visceral time} subdimension describes whether a visualization relies on the user's experience of time passing while navigating.

\begin{wrapfigure}[\iconlines]{l}{\iconwidthouter\linewidth}
  \centering
  \includegraphics[width=\iconwidthinner\linewidth]{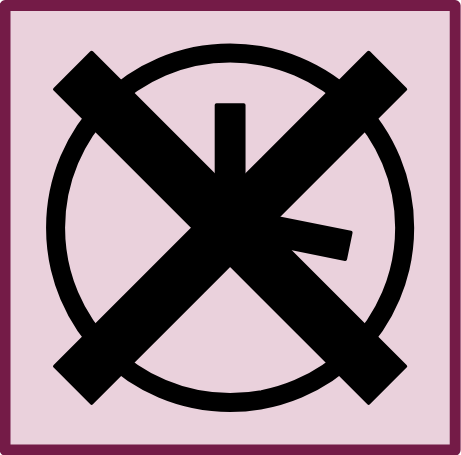}
\end{wrapfigure}
\noindent{\textit{No.}} Most of our corpus examples do not rely on the experience of time. For example, in Rivet, the user navigates by clicking to instantly zoom, meaning that users can fully navigate through the visualization very quickly. The decision to focus on rapid navigation over the use of visceral time makes sense in most cases, as arbitrarily slowing down visualization use could leave users frustrated and make tools less efficient.

\begin{wrapfigure}[\iconlines]{l}{\iconwidthouter\linewidth}
  \centering
  \vspace{-10pt}
  \includegraphics[width=\iconwidthinner\linewidth]{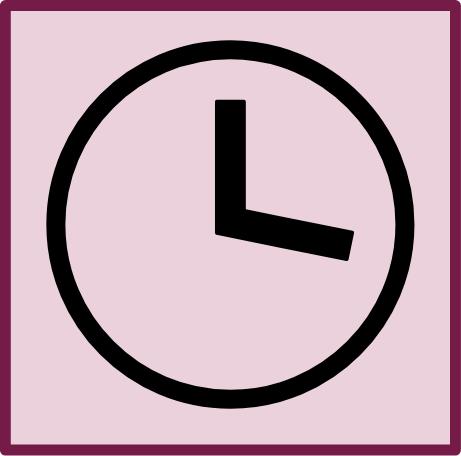}
\end{wrapfigure}
\noindent{\textit{Yes.}} Some visualizations of fully unfamiliar datasets do rely on \textbf{visceral time}. In Powers of Ten, video playback at a standard speed takes nearly ten minutes, so the sense of the time required to zoom at a constant rate between the scales aids the watcher's conceptualization of the difference between the scales by reinforcing the amount of zooming required to navigate between scales.

The Trail of Time, a large physical timeline that people hike along in the Grand Canyon where each meter represents one million years \cite{karlstrom2008trailoftime}, is an example of using the significant amount of time it takes for a visitor to complete the hike to help them conceptualize the multi-billion-year timeline. Similarly, the Deep Sea, a visualization of the depth of the sea which relies on users scrolling down from the water's surface to the seafloor \cite{agarwal2019thedeepsea}, also uses the significant amount of time required to complete this interaction to communicate extreme depth. While we do not include these examples in our design space as they are not examples of multiscale visualization, we mention them here as they are clear examples of the use of visceral time in visualization.

\subsection{Familiarity}

The \textbf{familiarity} dimension has a single subdimension: \textit{concrete}. It describes whether visualizations \change{use metaphors or analogies; that is, whether they} compare familiar objects to unfamiliar one\change{s} to help users understand the scale of the unfamiliar ones, and is related to the concept of concrete scales \cite{chevalier2013using}. Examples of familiar objects from our corpus are meters, days, and dollars.

\begin{wrapfigure}[\iconlines]{l}{\iconwidthouter\linewidth}
  \centering
  \vspace{-10pt}
  \includegraphics[width=\iconwidthinner\linewidth]{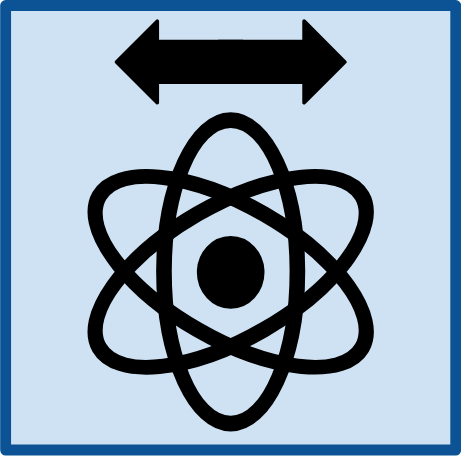}
\end{wrapfigure}
\noindent{\textit{No.}} Rivet is an example of a visualization that does not incorporate familiar objects as all items in the dataset are events in a trace log which occur during time spans that are much smaller than those humans are familiar with.

\begin{wrapfigure}[\iconlines]{l}{\iconwidthouter\linewidth}
  \centering
  \vspace{-10pt}
  \includegraphics[width=\iconwidthinner\linewidth]{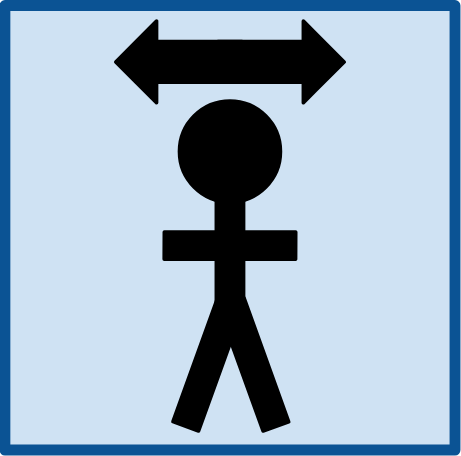}
\end{wrapfigure}
\noindent{\textit{Yes.}} Powers of Ten begins focused on familiar object of a picnic blanket near a meter in size but zooms to show very small or very large objects which have sizes that humans do not directly interact with.

\section{Strategies}
\label{sec:strategies}

\begin{figure}
    \centering
    \includegraphics[width=\linewidth]{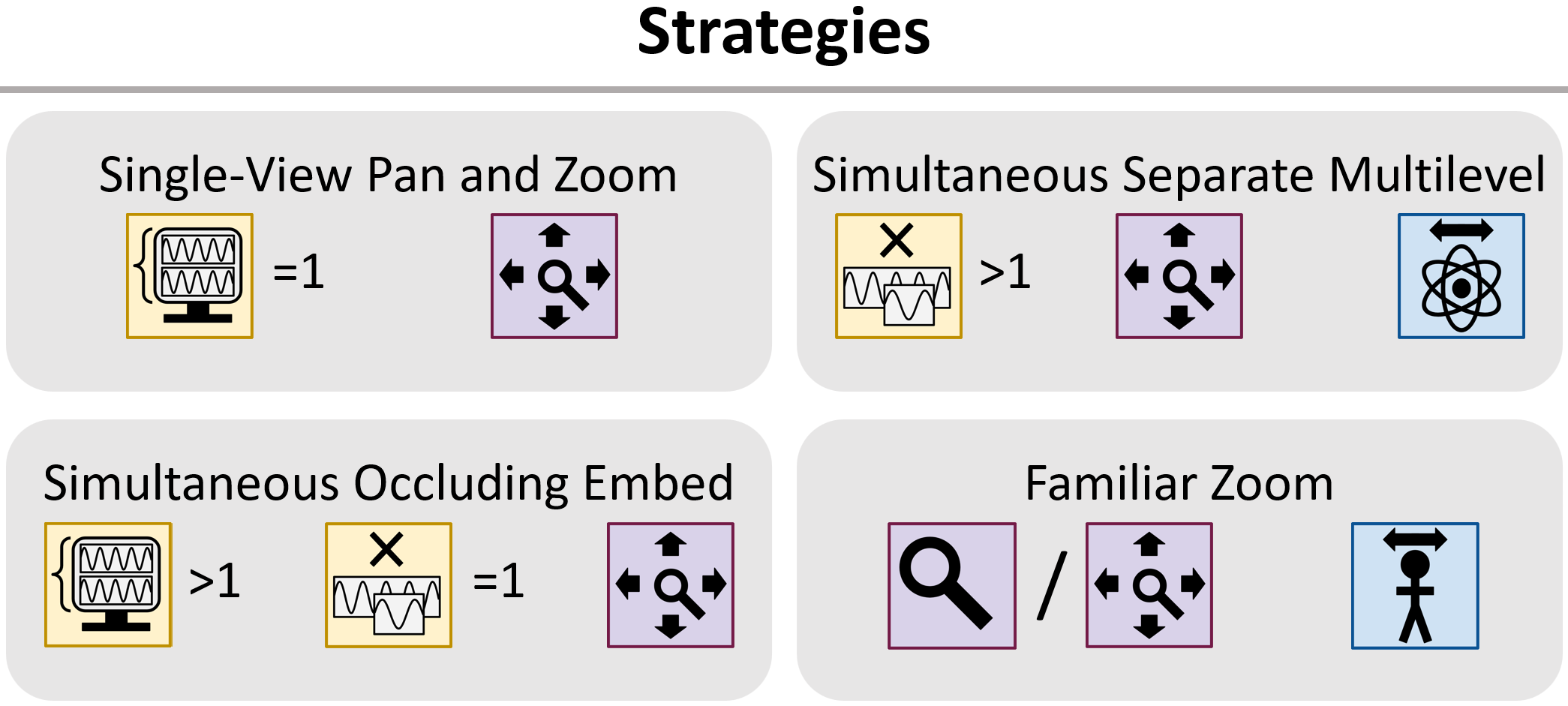}
    \caption{The four strategies for designing multiscale visualizations.}
    \label{fig:strategies}
\end{figure}

We identified four high-level strategies for designing multiscale visualizations which arose in a second-stage analysis from our iterative coding of our example corpus with respect to their dimension choices: Single-View Pan and Zoom, Simultaneous Occluding Embed, Simultaneous Separate Multilevel, and Familiar Zoom. The strategies are both concise, in that they are simple to reason about due to their lack of complexity, and disjoint, in that they do not overlap. While each strategy includes a set of dimensions which must use certain choices, other choices, unmentioned below, are unrestricted. These restricted choices only come from three of the subdimensions, indicating that these subdimensions are the most central to the design space. The strategies are shown in Figure~\ref{fig:strategies}. 

\change{The strategies are intended to represent common existing design patterns, not to propose novel ideas. Unsurprisingly, many of them are related to general strategies in visualization, such as Overview+Detail or Focus+Context \cite{cockburn2009review}. Below we discuss which of our identified strategies overlap with these general ones.} We now describe and discuss each strategy, primarily referring to our representative examples illustrated in~\ref{fig:example-superfig} but also referring to other interesting examples from the corpus to support our discussion.

\subsection{Single-View Pan and Zoom}

20 of our corpus examples use the Single-View Pan and Zoom strategy, which involves multiple \textit{total} scales but only one \textit{simultaneous} scale and \textit{both} zooming and panning. Given the large number of examples using this strategy, there are many variations. While all of the examples which use this strategy have only a single simultaneous scale, the number of total scales can vary significantly depending on the number of scales or levels within the data. 

Some examples that use this strategy rely on pointing and scrolling with the mouse, while others allow the user to click and drag to choose the next viewing window. Others force the user to choose a discrete zoom option from a set, rather than allowing continuous zooming, such as Large Viewing Vis, which also stands out within this group for its use of \textit{physical} navigation, although the overall strategy is the same. 

Many of these visualizations use the same encodings at each scale, but some have different encodings at different scales, such as EVEVis.

\subsection{Simultaneous Occluding Embed}

\change{13} of our corpus examples use the Simultaneous Occluding Embed strategy, meaning that they have multiple \textit{simultaneous} scales but only one \textit{separate} scale as the \textit{simultaneous} scales occlude each other in some way. They also need \textit{both} zooming and panning. 

Many of these examples use the inset zoom\change{, an example of the Overview+Detail pattern reminiscent of a video game minimap,} or lens zoom techniques \cite{tominski2017interactive}, \change{an example of the Focus+Context pattern} where a zoomed-in area appears in a window on of a visualization. This strategy limits the zoomed-in window to be smaller than the rest of the \change{view}, as full occlusion would result in only a single \textit{simultaneous} scale. \delete{Sometimes} \change{This} window \change{can occlude} the area being zoomed into, like in Melange. In other examples, such as FingerGlass \cite{kaser2011fingerglass}, the window occludes a separate part of the \change{view}, sometimes chosen by the user.

Similar to the Single-View Pan and Zoom strategy, \textbf{step size type} can vary. However, our corpus does not include any examples of this strategy where the user must use \textit{physical} navigation. Many of these examples use \textit{marks} to show \textbf{association} between the scales, but three include no \textbf{association}. Multi-Foci COVID Vis, a visualization of COVID data in selected geographic areas \cite{mactavish2022multifocicovidvis}, is the only example using this strategy to also use \textit{different} encodings on different scales.

\subsection{Simultaneous Separate Multilevel}

10 of our corpus examples use the Simultaneous Separate Multilevel strategy\change{, a subset of the Overview+Detail pattern}, which describes visualizations where there are multiple \textit{separate} scales, meaning also that there are multiple \textit{simultaneous} and \textit{total} scales, but which do not rely on \textbf{familiarity}. The difference between this strategy and Simultaneous Occluding Embed is that multiple scales appear without occluding one another, meaning that all scales can be of equal size.

Similar to both Single-View Pan and Zoom and Simultaneous Occluding Embed, examples using this strategy can both use \textit{different} encodings or the \textit{same} encodings on the varying scales. We find that all options for \textbf{association} are in use by at least one example in this group.

We found that when encoding choices across the separate scales of examples which use this strategy are the same, the different scales are aligned and either stacked on top of each other or placed beside each other with most of them using some form of association between them. In contrast, when the encoding choices are different, the scales are unaligned and placed in different completely separate views with no association.

All but one example that used this strategy employed \textit{both} panning and zooming, and all of these examples with navigation relied on \textit{digital} navigation. The one example that did not use \textit{both} panning and zooming was Temp Earth \cite{bredenberg2012tempearth}, which divides the axis into pieces, each of which has a different scale which is a multiple smaller than the one on its left.

\subsection{Familiar Zoom}

9 of our corpus examples use the Familiar Zoom strategy, which relies on \textit{zooming} through a series of scales that include at least one familiar scale. All Familiar Zoom examples must also incorporate \textbf{familiarity}. We also categorize an example that has no navigation but shows multiple \textit{separate} scales with \textbf{association} by \textit{marks} between them, xkcd Money \cite{munroe2011xkcdmoney}, as Familiar Zoom. Some of these examples focus on helping users conceptualize large scales, some with small scales, and some with both.

These examples rely on the user making comparisons between the different scales with varying \textbf{step size type}, beginning with a familiar scale. \textit{Familiarity} is only used in this strategy as the focus of these examples was to help users understand the concept of scale, rather than supporting analysis of data. In contrast, users conducting analysis tasks often have domain expertise and do not depend on the use of familiar objects or scales, which could clutter the visualization without providing any benefits. 

Most of these examples only show one \textit{simultaneous} scale, with the zooming modifying the scale of the single view. However, three of these examples have multiple \textit{separate} scales. Two of the examples with multiple \textit{separate} scales include a form of \textbf{association} between the scales, likely to aid the comparison between them. While most of these examples do not rely on \textbf{visceral time}, Powers of Ten \cite{eames1968powersoften} does.

\section{Missed Opportunities}
\label{sec:missed-opportunities}

We discuss benefits and limitations of choices within dimensions and how some examples may have benefited from alternate choices, going into detail for one example for each choice as case studies.

\subsection{Employing Physical Navigation}

Dynamic navigation is well-used in multiscale visualizations. Navigation is useful for analysis in these visualizations as it allows the user to choose regions to analyze in greater detail. We also see it used in examples intended for presentation, possibly to increase engagement through interactivity or to avoid overwhelming users by gradually revealing information. Most the corpus examples were digital and had to use input \textit{devices} to control navigation, but a few examples relied on physical movement. Ball et al.~found that physical navigation is beneficial and preferred by users in the right conditions \cite{ball2007move}, so we encourage design teams with sufficient resources, particularly space and materials, to consider employing some use of physical movement\delete{ by the user}.

\noindent{\textbf{Case Study:}} One example which could have benefited from physical navigation is xkcd Money \cite{munroe2011xkcdmoney}. While our stylized version in Figure~\ref{fig:example-superfig}\subfigxkcd is relatively simple, the real version has significant detail at each zoom level. As it is hosted online, users must digitally zoom in to the levels to see the finest details: tiny individual squares each representing a unit of money. This navigation is necessary to fully appreciate the example, but it being digital leads to challenges, as it can be disorienting. If xkcd Money were drawn onto a large wall, similar to Large Viewing Vis \cite{isenberg2013largeviewingvis}, users could navigate it more smoothly and maintain a sense of the full image using their peripheral vision even while physically zoomed in to see details.

\subsection{Simultaneous and Separate Scales}

The majority of the corpus examples used small \textit{total}, \textit{simultaneous}, and \textit{separate} \textbf{counts}. Using a single \textit{separate} scale or a small number of \textit{separate} scales can allow for more space to be dedicated to detail within the scale, reduce the complexity of the visual representation, and reduce the complexity of the navigation required. However, increasing the number of \textit{total} scales that the user can navigate to can help to show more detail, and increasing the number that are \textit{simultaneous} and \textit{separate} can be beneficial for comparison across scales or drill-down navigation. Here is Today \cite{twyman2020hereistoday} may have benefited from the use of multiple simultaneously visible scales, as its navigation between scales was sometimes hard to follow due to multiple changes of direction in the direction of motion of the timeline.

\begin{figure}[ht]
    \centering
    \includegraphics[width=.7\linewidth]{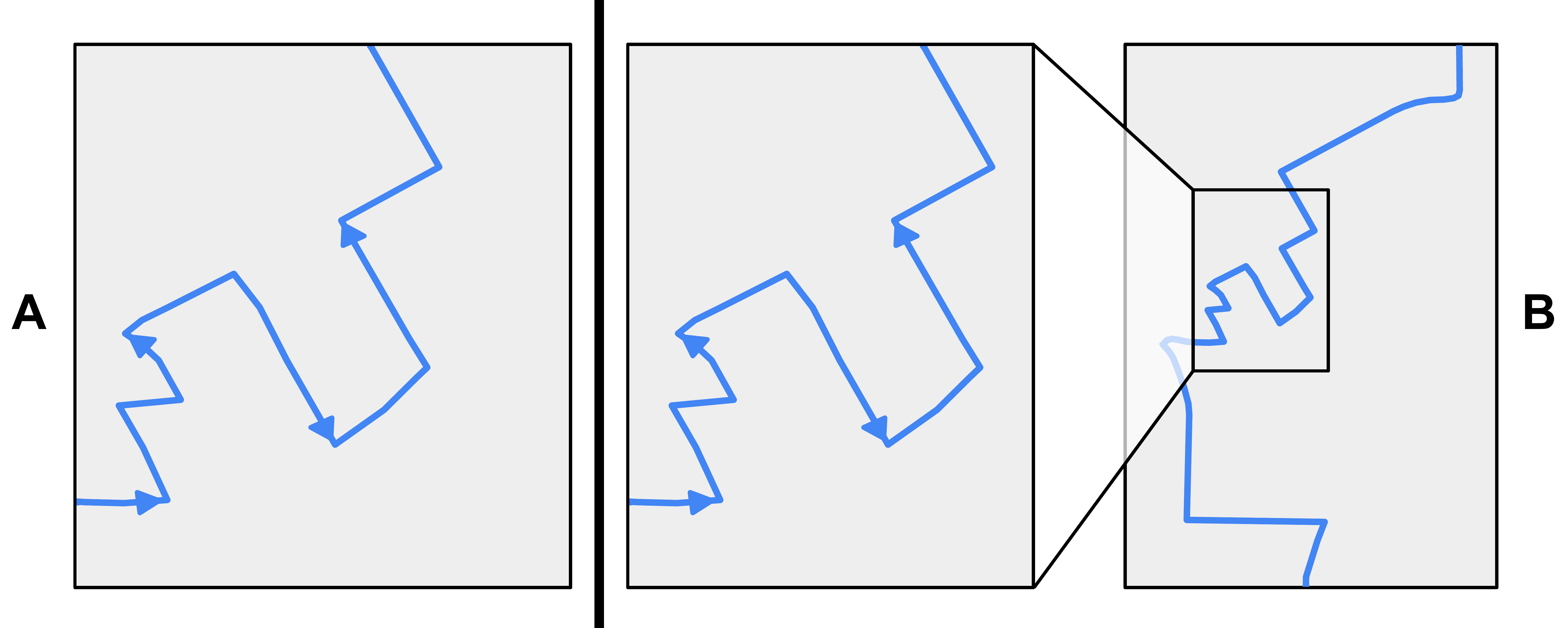}
    \caption{\change{Proposed} redesign of Hierarchical Route Maps \cite{wang2014hierarchicalroutemaps}. A) \change{Stylized} original. B) \change{Stylized} redesign, with \delete{a} second \textit{separate} scale showing an overview connected using association by marks.}
    \label{fig:casestudy-hierarchicalroutemaps}
\end{figure}

\noindent{\textbf{Case Study:}} Many of the examples using the Single-View Pan and Zoom strategy may have benefited from additional \textit{separate} scales, as they can support the user in navigation by allowing the user to skip to a different point of focus without needing to pan around or zoom out and back in. One such example is Hierarchical Route Maps, a visualization of navigational routes with zooming and panning capabilities to help navigators\delete{,} such as drivers or cyclists\delete{, in both dense and sparse areas} \cite{wang2014hierarchicalroutemaps}, shown in stylized form in Figure~\ref{fig:casestudy-hierarchicalroutemaps}A. In this corpus example, the tool automatically zooms in when a user is navigating an area with detailed route information. However, when it zooms in, the user may still need the zoomed-out version of the map, for example to orient themselves using landmarks only visible on the zoomed-out version. To alleviate this, Hierarchical Route Maps could have included a view showing the higher zoom level at all times, and only zoomed in in a second view, which would allow for both zoom levels to be visible at once. It could have also incorporated association between these scales to further orient the user. This hypothetical version of the tool is shown in Figure~\ref{fig:casestudy-hierarchicalroutemaps}B.

\subsection{Use of Differing Scale Encodings}

Most examples with multiple scales use the \textit{same} visual encoding on each scale. In our corpus, most examples which use \textit{different} encodings on different scales have pre-existing, real-world structure. However, we believe this design choice may also be beneficial for other scenarios. If the visualization encodes large quantities of data across multiple scales, then some scales will likely have much more data to encode than others, which may change what encoding is most effective. Multiscale Trace \cite{ezzati2014multiscaletrace} uses \textit{different} encodings on different scales for this reason.

In a similar scenario, higher-level scales may be used simply for navigation, to find smaller regions to analyze in more detail. In this case, the user is using different scales for different tasks, and the designer should consider this difference when making visual encoding choices for the different scales. Europe OSM \cite{zacharopoulou2021europeosm} uses \textit{different} encodings on different scales for this reason.

\noindent{\textbf{Case Study:}} An example where using \textit{different} encodings on different scales may have been beneficial is Rivet \cite{bosch2000rivet}. Our simplified version of Rivet's multi-tier strip chart shows a sine wave, but the real version is a dense visualization of computer systems data. It supports users in conducting analysis on details in the data by allowing them to navigate via zooming in multiple times, meaning that users use the higher levels of zoom for identifying regions to analyze and the lower levels of zoom to actually conduct analysis. This difference in tasks on different zoom levels suggests that different encoding choices may have been beneficial. For example, the designers could have chosen visual encodings for the higher levels that are better for surfacing features for analysis.

\subsection{Explicit Association Between Scales}

While \textbf{association} by \textit{marks} or \textit{channels} is not underused in our corpus, examples which did not use it may have benefited from it. Using \textbf{association} is helpful for comparison across multiple scales as it can explicitly show a change in mapping by relating the same item across scales. However, association is also helpful for user tasks other than across-scale comparison. For example, if the user is intended to use the multiple scales to navigate to a smaller region for analysis, using association can help the user keep track of the zoomed-in location on the larger-scale landscape. 

\begin{figure}[ht]
    \centering
    \includegraphics[width=0.7\linewidth]{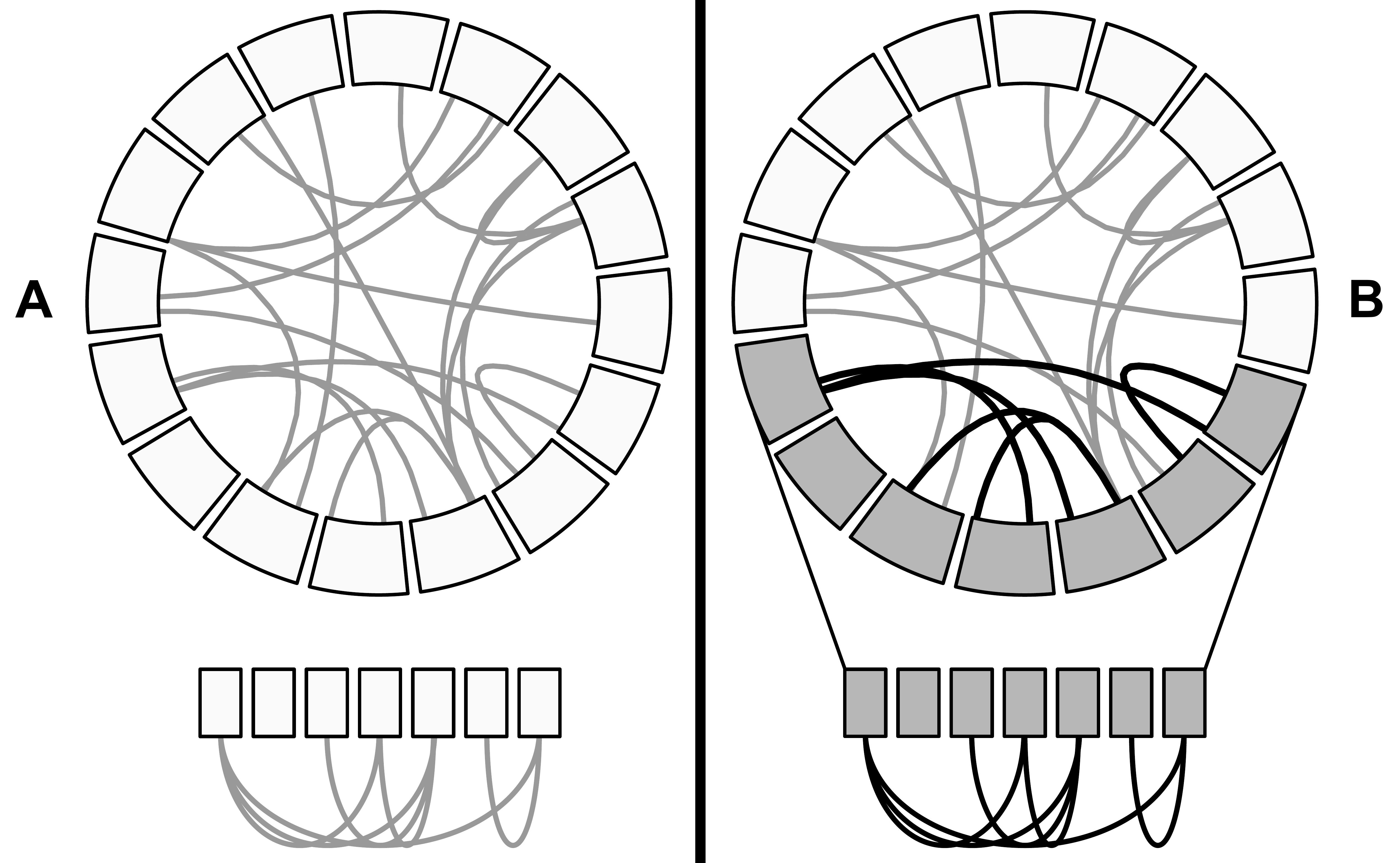}
    \caption{A proposed redesign of Chromoscope \cite{lyi2023chromoscope}. A) \change{Stylized} original. B) \change{Stylized} redesign, \change{with} association by \textit{both} marks and channels\change{.}}
    \label{fig:casestudy-chromoscope}
\end{figure}

\noindent{\textbf{Case Study:}} One example that uses the Simultaneous Separate Multilevel strategy, Chromoscope \cite{lyi2023chromoscope}, has two separate scales showing genomic data but lacks any \textbf{association} between those scales. The higher-level scale is a radially aligned and shows many different attributes of the data, including lines connecting start and end points of structural variants on a gene. It also has a separate scale below which is linearly aligned, which shows an individual structural variant selected by the user. We show a stylized version of this example in Figure~\ref{fig:casestudy-chromoscope}A. The lack of association between these two scales makes it challenging to see which structural variant has been selected, which could be disorienting if the user needs to look back and forth between the scales. The designers could have chosen to use either or both marks and channels to show association between these scales. For example, lines could be used to connect the selected structural variant on the radial scale to the linear scale, or a highlight colour could be used on the selected structural variant in both scales. A stylized version of our proposed redesign, which incorporates both the lines and the highlight colour, is shown in Figure~\ref{fig:casestudy-chromoscope}B.

\section{Visualization Context: Analysis vs. Presentation}
\label{sec:visualization-context}

Through our in-depth work with the example corpus and the design space, we observed some correlation between design choices, strategy choice, and source and the context that a visualization was used in. The most obvious pattern is the use of the Familiar Zoom strategy for examples used for presentation: the communication of already-gleaned insights. This pattern arises due to the common use of \textbf{familiarity}, a design choice only included in the Familiar Zoom strategy, within presentation-focused examples. For analysis-focused examples, where the intention is to glean new insights, users are often expected to be domain experts, leading to less need for explicit \textbf{familiarity}. Similarly, visceral time is only used for presentation as intentionally slowing down interaction time makes analysis less efficient.

After observing these differences, we decided to further analyze the coded corpus for other differences. The use of \textit{different} \textbf{encodings} on different scales is much more common for analysis-focused visualizations. We conjecture that this is often due to real-world structures which are represented differently in the data and the need to complete different tasks at different scales. Presentation-focused visualizations tend to have simpler or more consistently-structured datasets as they are intended to communicate insights without requiring an understanding of any real-world structures. We believe that designers of presentation-focused visualizations should consider using \textit{different} \textbf{encodings} on different scales, as it could emphasize the differences in scale and allow for different tasks to be emphasized to viewers. 

Presentation-focused visualizations tend to be simpler in other ways too. When visualizations incorporated \textit{both} \textit{panning} and \textit{zooming}, we found that they were intended to have open-ended navigation, where the navigation is driven fully by the user, typically to support analysis. When only \textit{zooming} was in use, the designs had an intended path to follow through the data, either by panning along a large scale or by zooming along many scales. Similarly, visualizations made for presentation were less likely to have user controlled navigation, leading to a more guided experience. \change{These presentation-focused visualizations follow author-driven narrative patterns like the Interactive Slideshow or the stem of the Martini Glass Structure~\cite{segel2010narrative}.}

Finally, the majority of presentation-focused examples we found were used in practice rather than studied in academic literature. Future work could investigate the use of multiscale visualization in this context to better understand the impact of the design choices and strategies outlined in this work.

\section{Discussion}
\label{sec:strengths-and-limitations}

\change{We assess the descriptive, generative, and evaluative power of 
our design space and strategies, and discuss limitations and future work.}


\change{As is common in design space papers, we provide a preliminary evaluation of the design space through the lens of descriptive, generative, and evaluative power~\cite{beaudouin2004designing}.} The design space has strong descriptive power\change{: all} meaningful differences from our analysis are distinguishable. \change{All included dimensions are required to make examples sufficiently distinguishable; excluding a dimension would lead to inadequate descriptive power.} We have further confidence in its completeness and descriptive power because we have evidence of saturation: no example in the systematic set, which we found and coded during the final Refine stage, required any changes to the design space to describe. 
(Although we further reflected and refined the dimensions, choices, and strategies after adding the systematic set, we were able to fully describe all examples in the systematic set with the design space prior to the final modifications.) The four strategies also demonstrate the descriptive power of the design space, because it is defined in terms of choices within the design space dimensions\change{, and our final set of dimensions are necessary to identify strategies}. The strategies themselves also have descriptive power, in that they provide a disjoint partition of the set of examples. One limitation of our work is that the set of strategies may not be complete: although they do fully describe our example corpus, future designs may use new strategies.

The design space also has generative power. Analysing it revealed missed opportunities within our example corpus which, if available during their design, could have resulted in changes to some examples. In the future, it could inform designers about design possibilities that they may not have considered without this specific prompting. Our strategies also demonstrate strong generative power, because the set of four strategies is a very concise set of options. Choosing one of these strategies can inform and speed design by immediately constraining some of the design choices. We note that the concise set of just four strategies may be useful for quickly choosing which  out-of-the-box solutions to use when novelty is not required; in contrast, the more detailed design space may be useful for generating custom visualizations.


One limitation of this paper is that we have not yet validated this design space or strategies in terms of evaluative power, an effort we leave for future work. 

Another limitation of the design space and strategies is that our collection of examples from real-world use for the corpus was opportunistic rather than systematic. While we did systematically search academic literature, a systematic search for practitioner examples would not be straightforward to conduct. In particular, finding examples used for education and communication by practitioners is challenging using visualization search terms; some are not even posted publicly online.

\section{Conclusion}

In this paper, we present a design space for multiscale visualization \change{with three dimensions and eight subdimensions, which we developed by iteratively coding a corpus of 52 examples}. \delete{The design space has three dimensions: Scales, Navigation, and Familiarity. These dimensions are split into eight subdimensions: count, step type, encodings, and association for Scales; type, mode, and visceral time for Navigation; and concrete for Familiarity.}
\delete{We collect and analyze a corpus of \change{52} examples from both research and practice, and code them according to the hierarchical dimensions of this design space to develop, validate, and illustrate the design space.} 
We \change{also contribute} a set of four high-level strategies for designing multiscale visualizations, \delete{which are shared approaches with respect to design space choices and are a partition such that each example fits into exactly one strategy.
We conduct} an analysis of missed opportunities for several examples\delete{ that considers alternative dimension choices and strategies, and include case studies to further clarify these ideas. We identify and discuss}\change{, and a discussion of} patterns in \delete{design space dimension and strategy choices}\change{multiscale visualization} in the differing \delete{visualization} contexts of analysis and presentation. \delete{Finally,} \change{We} evaluate the strengths and limitations of the design space, showing its descriptive and generative power.


\section*{Supplemental Materials}
\label{sec:supplemental_materials}

The supplemental material is available on OSF at \url{https://osf.io/wbrdm/}, released under a CC-BY-4.0 license. We provide an Excel file containing the example corpus, coded by the dimensions and by the strategies. In addition to the information shown in Table~\ref{tab:coded-examples-dimensions}, it contains the full titles of examples, the specific figure we coded for academic examples, and how we found the example. The coded design space is additionally available as an interactive website with filters at~\url{marasolen.github.io/multiscale-vis-ds/}.

\acknowledgments{
Our work was supported in part by NSERC DG RGPIN-2014-06309. We thank Laura Lukes and Nigar Sultana of the UBC Earth, Ocean and Atmospheric Sciences department, Francis Nguyen, Ryan Smith, and Steve Kasica of the UBC InfoVis Group, and Ben Shneiderman, for their helpful feedback.
}

\bibliographystyle{abbrv-doi-hyperref} 
\bibliography{vis25-2-references}    

\end{document}